**Coupling photogeneration with thermodynamic modeling of light-induced alloy segregation enables the discovery of stabilizing dopants**

Tong Zhu[1†], Sam Teale[1†], Luke Grater[1†], Eugenia S. Vasileiadou[2], Jonathan Sharir-Smith[1], Bin Chen[1], Mercouri G. Kanatzidis[2], Edward H. Sargent[1,2,3*]

[1]Department of Electrical and Computer Engineering, University of Toronto, 35 St George Street, Toronto, ON M5S 1A4, Canada

[2]Department of Chemistry, Northwestern University, 2145 Sheridan Rd, Evanston, IL 60208, USA

[3]Department of Electrical and Computer Engineering, Northwestern University, 2145 Sheridan Rd, Evanston, IL 60208, USA

*Correspondence and requests for materials should be addressed to E.H.S. (e-mail: ted.sargent@utoronto.ca).

[†]These authors contributed equally to this work.

**Abstract:** We developed a generalized model that considers light as energy contributed through the thermalization of excited carriers to generate excitation-intensity- and temperature-dependent phase diagrams. We find that the model replicates the light-induced phase segregation behavior of the MAPb(I,Br)$_3$ system. From there, we sought to study how best to design new, stable, mixed-halide alloys. The resultant first-principles predictions show that the pseudo halide anion BF$_4^-$ suppresses phase segregation in FA$_{0.83}$Cs$_{0.17}$Pb(I$_{0.6}$Br$_{0.4}$)$_3$, resulting in enhanced operating stability. The findings reveal that photostability is linked with the structure and electronic properties of materials and may be overcome using informed alloying strategies.

Mixed halide perovskites have rapidly gained attention for their use in efficient, low-cost solar cells and LEDs that benefit from bandgap tuning [1,2]. Reaching a desired bandgap is achieved by adjusting the ratio I/Br [3]; but this also leads to photo-induced instability, associated with halide segregation, when the Br fraction exceeds 20% [4]. Previous reports have used passivation and morphology control to manage halide segregation in alloys with < 30% Br content [5–9]; but no method has overcome the problem in alloys having greater Br content, such as in 40% Br perovskite films having the bandgap needed in perovskite tandem applications. These semiconductors suffer a 150 meV bandgap downshift under illumination, resulting in a ~15% drop in efficiency over the first few hours of operation [10].

Previous studies of photo-induced halide segregation suggest that it is caused by electric field gradients (polaron-induced lattice strain [11–13] or charge carrier gradients [14,15]), or is thermodynamic in origin [15–18]. Discriminating between these two theories is difficult as inconsistent results exist even from the experimental side (e.g. for the MAPb(I, Br)$_3$ system) [19] by considering different starting compositions and illumination conditions, and these are hard to be well understood. A microscopic model that generates accurate phase diagrams under illumination with a comprehensive picture of the free energy landscape remains an as-yet-unrealized goal for the field. [19,20] Herein we develop a model that constructs phase diagrams, predicting photo-induced halide segregation from first principles. We apply our phase diagrams to identify potential candidates that can stabilize a variety of perovskite alloys, even those with high Br content typically exhibiting pronounced segregation.

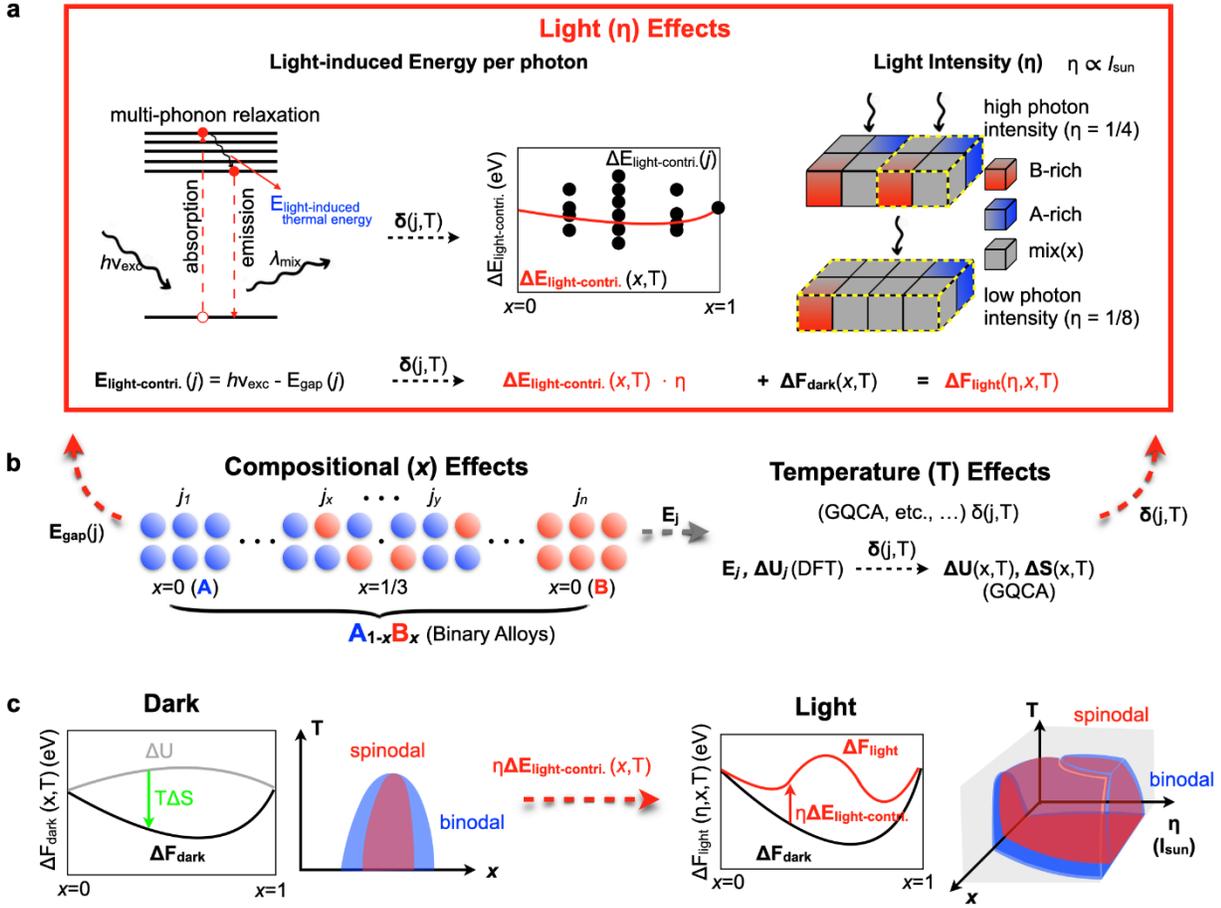

**FIG. 1. Introducing illumination in the generation of phase diagrams of alloyed semiconductors. a)** Schematic depicting the light-induced energy term and its intensity-dependence. **b)** Coupling with temperature and compositional effects. **c)** Phase diagram evolution when composition ($x$), temperature (T), and illumination intensity ($\eta$) are coupled.

Phase diagrams describe regions of compositional (in)stability for an alloy under an external stressor (temperature or pressure) and are crucial for alloy development [21]. These diagrams are often derived from first principles [22–25], enabling materials scientists to find methods to stabilize compositions. Thermodynamically, the full phase diagram of an alloyed material under dark conditions (no illumination) can be obtained from its Helmholtz free energy variation $\Delta F = \Delta U - T\Delta S$, where the energy of mixing, $\Delta U$, and the entropy of mixing, $\Delta S$, is determined through first-principles total energy calculations within the generalized quasi-chemical approximation (GQCA) [26] (Fig. 1b). This has been successfully employed in various alloy systems [27,28] including mixed halide perovskites. [17] To achieve this for any given alloy $A_{1-x}B_x$ ($x \in [0,1]$), the potential independent alloy configurations $j$ are first enumerated for each given composition $x$ (e.g., $x=0$: $j_1$; $x=1/3$: $j_x$, ..., $j_y$; $x=1$: $j_n$ – see Fig. 1b), and for each $j$ the total energy ($E_j$) is calculated from DFT. Then, the probability, $\delta(j, T)$, of finding the system in configuration $j$ for a given temperature T is obtained using the GQCA by utilizing the energy of mixing for each configuration ($\Delta U_j = E_j - [(1-x)E_A + xE_B]$), where the last two terms represent the fractions of the total energy of the parent compounds A ($x=0$) and B ($x=1$). Finally, utilizing the Boltzmann distribution, the mixing contribution to the alloy's internal energy $\Delta U$ and the configurational entropy $\Delta S$ as functions of composition $x$ and temperature T are obtained, and the Helmholtz free energy is evaluated as $\Delta F_{dark}(x, T) = \Delta U(x, T) - T \Delta S(x, T)$.

However, for the community to learn more about photoinstabilities within a material, it is important that phase diagrams be re-constructed such that light, temperature, composition, and phase stability are all coupled. What remains is to find an accurate description for the effects of illumination, $\eta \Delta E_{light-contri.}(x, T)$ (Fig. 1a), as a function of composition ($x$), temperature (T), and light intensity ($\eta$). Prior attempts posit that the excess energy of photo carriers is deposited in the perovskite lattice and this acts to drive instability. These attempts [16,18] share a central hypothesis: that the driving force is the result of photon funneling from wider to narrower bandgap regions [16,18], with the energy derived from experimentally determined bandgap differences between the parent mixed phase and Iodine-rich domains. However, as this experimentally determined term is not an appropriate description for each thermodynamic sampling configuration j, these models can still not offer a comprehensive picture of the free energy landscape and associated all coupled (including light, temperature, compositions) phase diagrams.

In this work, we hypothesize that the energy due to illumination can be treated within the thermodynamic framework through the above-bandgap photon energy deposited in the lattice due to thermalization via phonon relaxation. Then for each configuration $j$, this light contribution term can be defined as follows:

$$E_{\text{light-contri.}}(j) = \hbar v_{\text{exc}} - E_{\text{gap}}(j),$$

where the $v_{\text{exc}}$ is the frequency of the absorbed photon, and $E_{\text{gap}}(j)$ is the bandgap of configuration $j$ calculated from DFT.

By additionally considering the mixing contribution of this energy term, $\Delta E_{\text{light-contri.}}(j)$ based on the parent phases, we obtain:

$$\Delta E_{\text{light-contri.}}(j) = E_{\text{light-contri.}}(j) - [(1-x)\ E_{\text{light-contri.}}(A) + x\ E_{\text{light-contri.}}(B)]$$

$$= -\ [E_{\text{gap}}(j) - ((1-x)\ E_{\text{gap}}(A) + x\ E_{\text{gap}}(B))]$$

$$= -\ \Delta E_{\text{gap}}(j),$$

which only negatively depends on the mixing contribution of the bandgap energy, $\Delta E_{\text{gap}}(j)$ (defined as the band gap difference between the actual configuration, $j$, and the fractions of the parent compound A ($x=0$) and B ($x=1$)). It is otherwise independent of photon energy. Through this assumption, we successfully find an explicit expression for the light contribution of any specific thermodynamic sampling configuration $j$.

By further utilizing the probability $\delta(j, T)$ given by the GQCA, we obtain $\Delta E_{\text{light-contri.}}(x, T)$ which is coupled with the composition effect ($x$) and temperature effect (T) using the Boltzmann distribution (Fig. 1a). The combination of $\Delta E_{\text{light-contri.}}(x, T)$ and the $\Delta F_{\text{dark}}(x, T)$ leads to the excitation-intensity-dependent expression shown in Fig. 1a. Light intensity in this model is described through the photon deposited energy density, $\eta$: the ratio of absorbed photon energy units ($E_{\text{light-contri}}$) to unit cells under illumination. Thus, the excitation intensity-dependent system free energy is:

$$\Delta F_{\text{light}}(\eta, x, T) = \Delta F_{\text{dark}}(x, T) + \eta\, \Delta E_{\text{light-contri.}}(x, T).$$

The phase diagram (Fig. 1c) evolves from a 2D diagram (based on composition and temperature) to a 3D diagram that includes illumination intensity. Our hypothesis of energy dissipation in perovskites is not based on photon funneling, but rather on the absorption and subsequent dissipation of photon energy within the material, through avenues such as ion migration. As a result, the correlation between energy dissipation and photocarrier lifetime is not direct. However, a rough equivalence between light intensity, $\eta$, and the photon flux, $\Phi_{photon}$, can be estimated (further details in Supplemental Materials Text 1). Thus we estimate that 1 sun illumination intensity for the MAPb(I, Br)$_3$ system (based on bandgap of 40% Br, ~1.75 eV) is roughly equivalent to $\eta = 1/6$, while for the Cs$_{0.25}$FA$_{0.75}$Pb(I, Br)$_3$ system (based on bandgap 40% Br, ~1.78 eV) 1 sun intensity corresponds to $\eta = 1/7$.

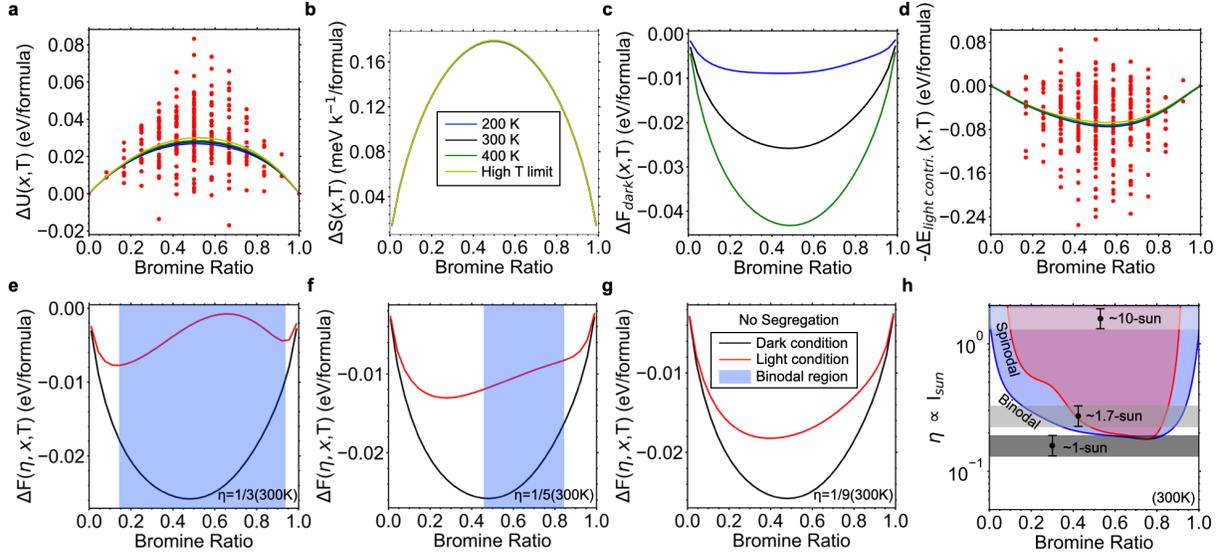

**Fig. 2. Benchmarking the model investigated herein in the case of MAPb(I$_{1-x}$Br$_x$)$_3$. a)** The energy of mixing $\Delta U(x, T)$, the red markers are the values calculated for each configuration ($\Delta U_j$ = $E_j - [(1-x)E_A + xE_B]$). The solid lines show the behavior for the alloy at 200 K (blue), 300 K (black), 400 K (green), and for a completely random alloy in the high T limit (yellow) within the GQCA. **b)** the entropy of mixing $\Delta S(x, T)$, **c)** the Helmholtz free energy variation in the dark condition $\Delta F_{dark}(x, T)$, and **d)** the light contribution terms $\Delta E_{light\text{-}contri.}(x, T)$ as functions of the MAPb(I$_{1-x}$Br$_x$)$_3$ alloy composition $x$ and temperature T. The red dots in $\Delta U(x, T)$ and $\Delta E_{light\text{-}contri.}(x, T)$ are the values calculated for each configuration $j$ from density functional theory (total energies in $\Delta U(x, T)$, and band gaps in $\Delta E_{light\text{-}contri.}(x, T)$). The solid lines show the behavior of the alloy at 200 K (blue), 300 K (black), 400 K (green), and for a completely random alloy in the high T limit (yellow) within the generalized quasi-chemical approximation (GQCA). **e-g)** The excitation intensity-dependent free energy variation for MAPb(I$_{1-x}$Br$_x$)$_3$ at 300 K with different light intensities ($\eta$=1/3,1/5,1/9). The solid lines show the behavior of the alloy in dark conditions (black) and illuminated conditions (red). The predicted phase segregation regions (miscibility gaps/binodal regions) are shown as a blue shaded area. More detailed free energy variation plots for different $\eta$ are in Fig. S4. **h)** The predicted phase diagram along with different light intensities at 300K. The binodal and spinodal lines and regions are shown in blue and red, respectively. The light intensities are illustrated by greyscale shaded areas (~1-sun, 1.7-sun, 10-sun). The detailed interpretation of $\eta$ to the experimentally used light intensity is in the supporting information Text 1. The temperature effects on this phase diagram can be found in Fig. S5.

We used MAPb(I$_{1-x}$Br$_x$)$_3$ based on a tetragonal structure to benchmark the model. First, we calculated the enthalpy and entropy of mixing (Fig 2 a,b, respectively) and used this to calculate the system's free energy under dark conditions (Fig. 2c). $\Delta F_{dark}(x, T)$ is virtually symmetric about $x$ = 0.5 at low temperature (200 K), a consequence of the symmetric distribution of the energy of mixing, $\Delta U_j$. Likewise, the convex profile of $\Delta F_{dark}$ at 300 K (Fig. 2c) means that a miscibility gap is not apparent (no points have the same tangent line), i.e. the alloy will not phase segregate: indeed, segregated domains remix in dark conditions after being exposed to light. [4] However, under illumination (Fig. 2e-g, Fig. S4a-f), $\Delta F(\eta, x, T)$ begins to show a concave shape for higher excitation intensity (correlated to $\eta$ = 1/5, 1/3, 1), i.e. the alloy will segregate as illumination is increased to ~ 1 sun ($\eta$ = 1/6, Fig. 2h). On the other hand, at lower excitation intensities ($\eta$ = 1/7, 1/9, 1/30), $\Delta F(\eta, x, T)$ closely resembles the case in dark conditions, showing no segregation and self-consistency of the model. (Fig. S4)

We found that the threshold below which Br content must be kept to avoid phase segregation is an intensity-dependent value (Fig. 2h). At ~1.7-sun at 300K, the transition from stable to unstable occurs when Br increases above ~ 15-34%, in agreement with experimental evidence [29] (~20% Br) at the same light intensity. Despite inconsistencies in reported experimental results [4,19,30,31] for $MAPb(I_{1-x}Br_x)_3$, this transition agrees well with different experiments as a function of composition and light intensity (Table S5), particularly when considering that this threshold is also affected by temperature (Fig. S5). We also note that the accuracy of this model at determining temperature contributions is restricted by the chosen supercell size (sampling configurations), and thus materials screening should be performed carefully.

In order to identify candidates with the potential to suppress light-induced phase segregation in a state of art wide-bandgap perovskite composition, $FA_{0.83}Cs_{0.17}Pb(I_{0.6}Br_{0.4})_3$ [32–34], we first conduct a global screening study across different A-cations ($APb(I_{1-x}Br_x)_3$, A = FA, Cs, MA, EA,

DMA, GUA), B-cations (CsB($I_{1-x}Br_x$)$_3$, B = Pb, Ba, Sn, Sr) and X-anions (FAPb($I_{1-x}X_x$)$_3$, X = Cl, BH$_4$, BF$_4$, CN, SCN) (Fig. S1-S3). Considering the computational cost, all perovskite structures are restricted to the cubic phase (the same as the FA$_{0.83}$Cs$_{0.17}$Pb(I$_{0.6}$Br$_{0.4}$)$_3$) based on 2×1×1 supercell (6 halides). As a result, we note that these predictions can be used to determine potential stabilization effects within a cubic phase framework, while the exact phase of the composition is reported to vary [35,36].

In the case of mixed (I, Br) perovskites, we found that using DMA or FA as the A-cation allowed for the highest non-segregating Br content ~ 50% ($\eta$=1/5, ~ 1 sun at 300K). While a pure FA film is not stable in the (cubic) perovskite phase, changing the A-site to FA-based, Cs stabilized compositions can enhance stability [35,37]. We also observed that the pseudo-anion BF$_4^-$ offers a large X-site tolerance, permitting a BF$_4^-$ concentration of 90% without phase segregation in mixed (I, BF$_4$) perovskites at $\eta$ = 1/5 (~ 1 sun), enabled by a deep valley in $F_{dark}(x, T)$ (Fig. S3) which corresponds to an increased enthalpy contribution due to mixing.

Due to this large X-site tolerance, we suppose that BF$_4^-$ may suppress light-induced phase segregation in the target (FA$_{0.83}$Cs$_{0.17}$Pb(I$_{0.6}$Br$_{0.4}$)$_3$). In order to minimize fluctuations in the bandgap, required for specific PV applications, we structure a refined study by considering the smallest BF$_4^-$ alloying ~8%, based on a 2×2×1 supercell (1 BF$_4^-$ in 12 halides) in FA$_{0.75}$Cs$_{0.25}$Pb(I,Br)$_3$ system (i.e., FA$_{0.75}$Cs$_{0.25}$Pb(I$_{0.92-x}$Br$_x$(BF$_4$)$_{0.08}$)$_3$ ). Unlike the system without BF$_4^-$ (Fig. 3a) in which a 40% Br composition segregates at intensities ranging from 1-sun to 10-suns, the 8% BF$_4^-$ alloyed system (Fig. 3b) will prevent a 40% Br composition from segregating at ~ 1-sun intensity while still segregating at 10-sun intensity. This implies that alloying the FA$_{0.75}$Cs$_{0.25}$Pb(I,Br)$_3$ system with BF$_4^-$ indeed suppresses light-induced segregation at operational light intensities (i.e. 1-sun) based on theoretical predictions.

To assess our findings experimentally, we fabricated thin films using the perovskite composition FA$_{0.83}$Cs$_{0.17}$Pb(I$_{0.6}$Br$_{0.4}$)$_3$ [32–34] and conducted photoluminescence (PL) measurements at 10-sun and 1-sun intensities to monitor halide segregation (Fig. 3c,e). Though films initially emitted ~700 nm light, a new emission emerged from an iodine-rich phase (~765 nm, ~12% Br, in Fig. 3b, matches with interpolated value in Fig. S15) after 10 minutes at 10 sun intensity (Fig. 3c), and after several hours at 1 sun intensity (Fig. 3e), consistent with prior studies [37], and in agreement with theoretical predictions. The composition of the segregated iodine-rich phase is also accurately predicted by the model (~13% Br at 10 sun, ~18% Br at 1 sun) when compared to the experimental findings (~12% Br for 1-sun and 10-sun). We also note apparent discrepancies (such as those in the 50% Br range, Fig. 3a) arise, which may potentially attribute to specific ion migration properties in this range and will not affect the overall predictive power of the model.

We then experimentally added 1% mol. FABF$_4$ to the perovskite precursor, limited by the solubility of FABF$_4$ and after finding that higher concentrations significantly lowered PL. We confirmed the incorporation of BF$_4^-$ through a combination of nuclear magnetic resonance, X-ray diffraction, absorption, photoluminescence, and photoelectron spectroscopy, in line with prior reports (more details in supplementary text 2, along with Fig. S16-S18) [37,38]. Samples excited using 1-sun intensity exhibited significantly suppressed segregation over a 12-hour period compared to the control (Fig. 3d,f). This strategy proved more effective than previously reported

Cl alloying [39] (Fig. S8, and Table S1). Reducing the Br concentration to 30% resulted in negligible segregation over a 12-hour period at 1-sun when $BF_4^-$ is added (Fig. S7a,b).

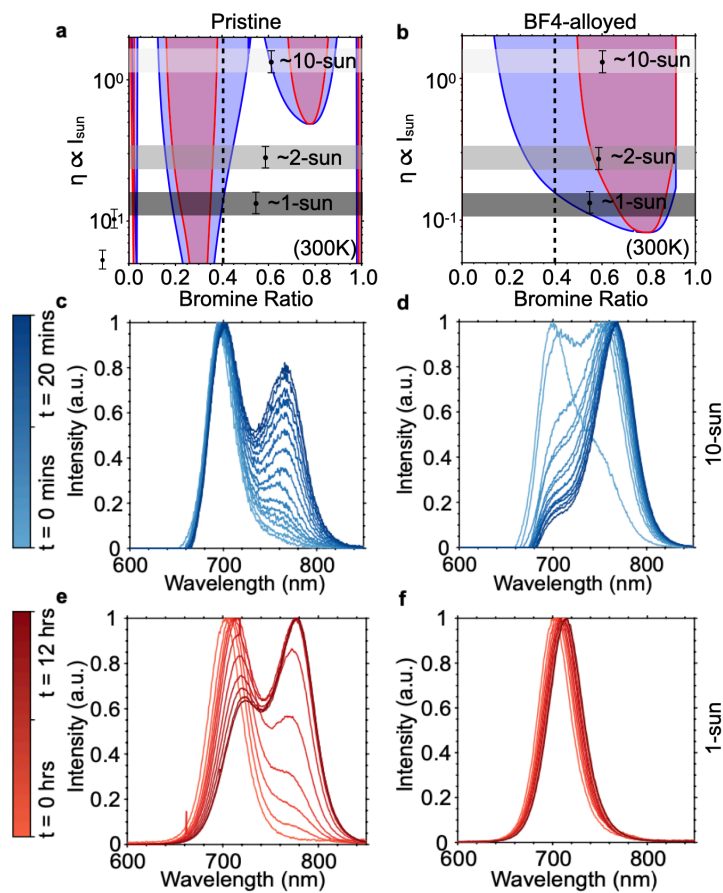

**Fig. 3: The role of $BF_4^-$ in increasing stability: computation and experiment. a)** Predicted phase diagram along with different light intensities at 300K for $FA_{0.75}Cs_{0.25}Pb(I_{1-x}Br_x)_3$ perovskite alongside PL tracking of $FA_{0.83}Cs_{0.17}Pb(I_{0.6}Br_{0.4})_3$ perovskite films at **c)** ~10-sun intensity for 20 minutes and **e)** ~1-sun intensity for 12 hours. **b)** A similar predicted phase diagram at 300 K is shown for $FA_{0.75}Cs_{0.25}Pb(I_{0.92-x}Br_x(BF_4)_{0.08})_3$ alongside PL tracking of $BF_4^-$ alloyed $FA_{0.83}Cs_{0.17}Pb(I_{0.6}Br_{0.4})_3$ films at **d)** ~10-sun intensity and **f)** ~1-sun intensity. The DFT predicted phase diagrams show binodal and spinodal lines and regions in blue and red, respectively. The estimated light intensities 1-sun, 2-sun, 10-sun are illustrated by greyscale shaded areas. With a Br ratio (40%), our experimental samples of interest are illustrated as a black dashed line. The interpretation of η to real light intensity can be found in the supporting materials Text 1. A more detailed Helmholtz free energy variation at 300K can be found in Fig. S6.

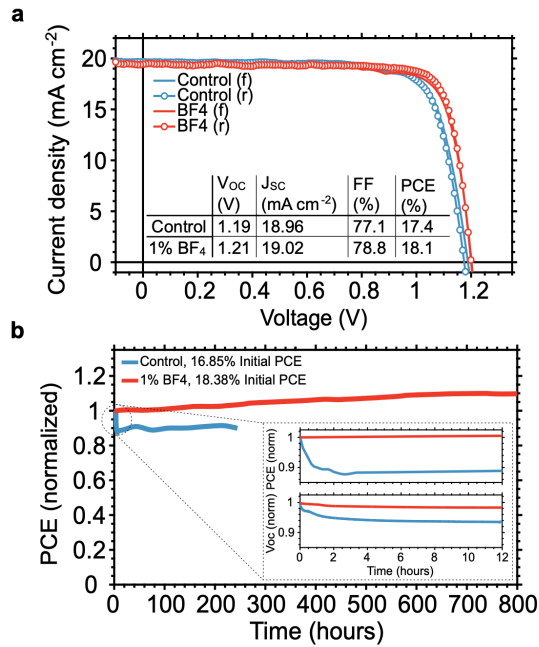

**Fig. 4. Solar cells employing BF$_4^-$. a)** *J-V* curves for control and BF$_4^-$ treated films, the inset table contains the average value from 15 devices of each variety. **b)** MPP tracking data from unencapsulated devices under 1-sun illumination over 800 hours, displays a zoom-in of the first 12 hours demonstrating a significant loss in PCE for the control film due largely to a drop in V$_{OC}$.

To further demonstrate the application of our model, we fabricated solar cells using Cs$_{0.17}$FA$_{0.83}$Pb(I$_{0.6}$Br$_{0.4}$)$_3$ with and without BF$_4^-$ alloying (detailed in methods and supplementary text 3). Under continual efficiency tracking, the BF$_4^-$ solar cell lost < 2% of its original open circuit voltage, V$_{OC}$, after 12 hours, compared to a 7% loss in the control. This resulted in the control dropping from 16.9% to 15.1% efficiency in the first 12 hours, while the BF$_4^-$ device retained over 18% efficiency for over 800 hours. This further validates the findings from our model, as halide segregation induces V$_{OC}$ loss over time as charges from Br-rich regions are funneled into narrow-bandgap I-rich domains losing energy in the process. [4] In addition, the loss in efficiency follows the same trend and timescale of the PL tracking experiments in Fig. 3c-3f [38,40] [4].

To summarize, due to photo-induced alloy segregation in mixed halide perovskites, it was revealed that most alloy compositions are inaccessible for practical applications. To better understand their behavior – and inform routes towards stability – phase diagram formalisms needed to be extended to accommodate light as an external stressor. We accomplish this using a generalized model, which incorporates illumination as energy deposited via the thermalization of excited carriers. Because the model is fully ab-initio, it can be used to provide a theoretical understanding of light-induced alloy segregation in any material that exhibits it. Additionally, further perturbations such as local strain, defects, or kinetic effects can be added to the model to further improve applicability and accuracy.

Using this model we generated illumination-dependent phase diagrams that accurately predict experimental behavior and used our model to stabilize a range of mixed halide alloys at operational light intensities (i.e. 1-sun). Our findings demonstrate that carrier-induced strain is not a prerequisite to describing photo instability. Instead, we reveal that this is an inherent property of mixed halide perovskite alloys, a byproduct of competition between fluctuations of the mixing contribution of the internal energy $\Delta U$, the configurational entropy $\Delta S$, and the fundamental gap $\Delta E_{gap}$, which can be overcome using informed alloying strategies.

**Data availability**
The data generated and/or analyzed during the current study are available from the corresponding author upon reasonable request.

**Code availability**
The codes and post-analysis/plotting tools for light effects and associated phase diagrams are available from https://github.com/ericzhut/Model-for-light-induced-phase-segregations

**Ethics declarations**

Competing interests
The authors declare no competing interests.

**Acknowledgments**

This research was made possible by the US Department of the Navy, Office of Naval Research Grant (N00014-20-1-2572 (EHS) and N00014-20-1-2725 (MGK)). This work was supported in part by the Ontario Research Fund-Research Excellence program (ORF7-Ministry of Research and Innovation, Ontario Research Fund-Research Excellence Round 7). ST would like to thank the Hatch Scholarship for supporting this work. Computations were performed on the Niagara supercomputer at the SciNet HPC Consortium. SciNet is funded by the Canada Foundation for Innovation; the Government of Ontario; Ontario Research Fund Research Excellence; and the University of Toronto.

**Supplementary Materials:**
Materials and Methods
Figure S1 – S19
Tables S1 – S5
Reference ( [4,29–31,40–52])

# Supplementary Information for

## Coupling photogeneration with thermodynamic modeling of light-induced alloy segregation enables the discovery of stabilizing dopants

Tong Zhu[1†], Sam Teale[1†], Luke Grater[1†], Eugenia S. Vasileiadou[2], Jonathan Sharir-Smith[1], Bin Chen[1], Mercouri G. Kanatzidis[2], Edward H. Sargent[1*]

1Department of Electrical and Computer Engineering, University of Toronto, 35 St George Street, Toronto, ON M5S 1A4, Canada

2[2]Department of Chemistry, Northwestern University, 2145 Sheridan Rd, Evanston, IL 60208, USA

*Correspondence to ted.sargent@utoronto.ca

**This PDF file includes:**

    Materials and Methods
    Supplementary Text
    Figs. S1 to S19
    Tables S1 to S5

**Materials and Methods**

Materials

All the materials were used as received without purification. Commercial ITO substrates (20 Ω/sq) with 25 mm x 25 mm dimension were purchased from TFD Inc. Lead iodide ($PbI_2$), Lead bromide ($PbBr_2$), cesium iodide (CsI), formamidinium hydroiodide (FAI), and formamidinium tetrafluoroborate ($FABF_4$) were purchased from TCI. $C_{60}$ (99.5%) and poly(methyl methacrylate) (PMMA), mean molecular weight 97,000 g/mol, were purchased from Sigma-Aldrich. Octylammonium bromide (OTABr) and formamidinium bromide (FABr) were purchased from Greatcell Solar. PCBM was purchased from nano-C. All the solvents used in the process were anhydrous and purchased from Sigma-Aldrich.

Methods

**Nickle oxide nanoparticle synthesis**

The $NiO_x$ nanoparticles (NCs) were prepared via the hydrolysis reaction of nickel nitrate referring to the previous work.[1,2] Briefly, 20 mmol Ni $(NO_3)_2$ · $6H_2O$ was dissolved in 20 mL of deionized water to obtain a dark green solution. Then, 4 mL of NaOH aqueous solution (10 mol $L^{−1}$) was slowly added into the solution while stirring. After being stirred for 20 min, the colloidal precipitation was thoroughly washed with deionized water three times and dried at 80°C for 6 hours. The obtained green powder was then calcined at 270°C for 2 hours to obtain a dark-black powder. The $NiO_x$ NCs ink was prepared by dispersing the obtained $NiO_x$ NCs in mixed solution of deionized water and IPA (3/1, v/v) with a concentration of 10 mg $mL^{−1}$.

**Solar cell fabrication**

ITO glass was cleaned by sequentially washing with detergent, deionized water, acetone and isopropanol (IPA). Before use, the ITO was cleaned with ultraviolet ozone for 20 min.

For inverted solar cells, the substrate was spin-coated with a thin layer of $NiO_x$ nanoparticle film from the $NiO_x$ NCs ink at 2,000 rpm for 30 s. The perovskite absorber layers were deposited inside the $N_2$-filled glove box with a controlled $H_2O$ and oxygen level to less than 1 ppm. The temperature inside was monitored to be 25-30 °C. The precursors for $Cs_{0.17}FA_{0.83}Pb(I_{0.6}Br_{0.4})_3$ perovskites were prepared by dissolving the $PbI_2$, $PbBr_2$, CsI, and FAI in DMF: DMSO (4:1 in volume) solvents.

For FABF$_4$ containing perovskites, 1 mole percent of FABF$_4$ was added to the precursor powders before solvating. Precursor solutions were formed by adding all precursor and additive powders to a 20 mL vial, then adding 1 mL of mixed solvent and then leaving the vial on a stirring hotplate set to 60°C for 30 minutes; all steps were undertaken in a glovebox. For the perovskite film fabrication, the substrate was spun at 2000 RPM for 30 s with an acceleration of 1000 rpm at first, and then at 6000 rpm for the 10 s with an acceleration of 6000 rpm/s. In the second step, 150 μL Anisole was dropped onto the substrate during the last 5 s of the spinning. The substrate was immediately placed on a hotplate and annealed at 100 °C for 20 min. For the surface treatment, the 2D solutions were prepared by dissolving the 2D ligand salt (OTABr) in chloroform. 1 mg mL$^{-1}$ of ligand salt was used, unless stated otherwise. The 2D layer was fabricated by depositing 150 μL of 2D solution onto the perovskite film surface, immediately after deposition the film was spun at a spin rate of 4,000 RPM for 30 seconds with a 4,000 RPM/s acceleration. The film was then annealed at 100 °C for 5 min. C$_{60}$/BCP or PCBM/ALD-SnO$_2$ were used as the electron transporting layer. C$_{60}$ was formed by evaporation, and deposition of ALD-SnO$_2$ was carried out in the PICOSUN R-200 Advanced ALD system. H$_2$O and TDMASn were used as oxygen and tin precursors. Precursor and substrate temperature was set to 75 °C and 85 °C, respectively. 90 SCCM N$_2$ was used as carrier gas. Pulse and purge times for H$_2$O were 1s and 5 s, respectively: and 1.6 s and 5 s for TDMASn. The total deposition cycle is 134, corresponding to 20 nm of SnO$_2$. PCBM was formed by spin-coating the PCBM solution (20 mg mL$^{-1}$ in chlorobenzene) at 1,000 rpm for 30 s and then annealed at 100 °C for 5 min. Then a thin and uniform BCP layer was deposited by drop-casting BCP dissolved in IPA in 2 to 3 drops while spinning the substrate at 5000 rpm. Finally, a 120 nm thick Ag contact was deposited on top of BCP or ALD-SnO$_2$ using thermal evaporation under high vacuum (<5 × 10$^{-7}$ Torr) in an Angstrom Engineering deposition system to produce a 0.053cm$^2$ active area cell.

**Perovskite film fabrication for thin film XRD, UV-VIS and PL measurements**
ITO glass was cleaned by sequentially washing with detergent, deionized water, acetone and isopropanol (IPA). Before use, the ITO was cleaned with ultraviolet ozone for 20 min. The Cs$_{0.17}$FA$_{0.83}$Pb(I$_{0.6}$Br$_{0.4}$)$_3$ or BF$_4^-$ containing Cs$_{0.17}$FA$_{0.83}$Pb(I$_{0.6}$Br$_{0.4}$)$_3$ perovskite was then prepared and deposited in the same manner as detailed in the above section for solar cell fabrication. After the perovskite films were annealed, a PMMA solution was prepared at a concentration of 150 mg/ml in chlorobenzene. 40 ul of the PMMA solution was then deposited dynamically on top of the perovskite layer at a speed of 2000 rpm for 30 seconds. The films were then annealed at 100°C for 5 minutes to drive off any residual solvent.

**Perovskite single crystal growth for characterization measurements**

$FAPbBr_3$ single crystals were grown via an inverse temperature crystallization method. Precursor powders, FABr and $PbBr_3$ were combined in a 20 mL vial to produce a 1.4M solution in a mixed gamma-butyrolactone, dimethyl formamide solution at a 1:1 volume ratio. For $FABF_4$ containing $FAPbBr_3$ single crystals, $FABF_4$ was added to vials containing the FABr and $PbBr_2$ precursor powders in respective (1%, 2%, 5%) mole percentages. After the precursor powders were completely dissolved, the solution was filtered with a 0.22 um pore size PTFE filter. The crystal growth process was then performed by placing the solution in a growth oven at room temperature and then increasing the growth temperature from 25°C to 80°C over a period of 24 hours, which corresponds to a rate of 2.3°C per hour. $FAPbBr_3$ single crystals were then obtained at the end of the growth cycle. Before any subsequent measurements were conducted on the $FAPbBr_3$ or $FAPbBr_3 + FABF_4$ single crystals, the single crystals were removed from their growth solution and washed x5 in acetonitrile to remove any residual surface $FABF_4$ and then left to dry in a vacuum oven for 12 hours.

**Device testing**

The current density-voltage (*J-V*) characteristics were measured using a Keithley 2400 source meter under illumination from a solar simulator (Newport, Class A) with a light intensity of 100 S5 mW cm$^{-2}$ (checked with a calibrated reference solar cell from Newport). *J-V* curves were measured in a nitrogen atmosphere with a scanning rate of 100 mVs$^{-1}$ (voltage step of 10 mV and delay time of 200 ms). The active area was determined by the aperture shade mask (0.049 cm$^2$ for small-area devices) placed in front of the solar cell. A spectral mismatch factor of 1 was used for all *J-V* measurements. The device testing chamber was left under ambient conditions for stabilized output measurements at maximum power point (MPP). Solar cells were fixed at the MPP voltage, (determined from *J-V* sweeps in both scanning directions) and current output was tracked over time.

**Stability testing**

Devices were placed in a homemade stability tracking station. Illumination source is a white light LED with intensity calibrated to match 1-sun conditions. For room temperature tests (ISOS-L-1I), device chamber was sealed and supplied with continuous $N_2$ purging. MPP were tracked by a perturb and observe algorithm that updates the MPP point every 10s.

**X-ray diffraction (XRD)**

XRD patterns were collected using a Rigaku D/Max-2500 diffractometer equipped with a Cu Kα1 radiation ($\lambda = 1.54056$ Å).

**X-ray photoelectron spectroscopy (XPS) characterization**

XPS measurements were conducted using a ThermoFisher Scientific (East Grinstead (UK)) K-Alpha machine using monochromatized Al Kα X-ray radiation. Samples were mounted on a steel plate and the takeoff angle was set to 90°. For surface characterization, a 400um spot size (2:1 ellipse, with the major axis being the noted spot size) was used. For sputtering XPS a 300um spot size was used. Survey spectra were acquired at a pass energy of 200eV. The corresponding point density on the energy axis was 1eV/step. These scans were performed in order to identify all the detectable species on the surface. Local scans were performed at high energy resolution, with a lower pass energy of 50 eV, and with correspondingly higher point density on the energy axis, with 0.1 eV between points. These spectra were used for quantification. The dwell time for the acquisition of these spectra was 50ms. Vertical sputtering/profiling was performed using 1keV $Ar^+$ ions, resulting in a 1500um crater size. Data were processed using Thermo Scientific Avantage software. Surface elemental compositions were calculated from background-subtracted (smart background) peak areas derived from transmission function corrected regional spectra. Sensitivity factors used to calculate the relative atomic percentages were provided by the instrument manufacturer.

**Solution nuclear magnetic resonance (NMR)**

All NMR data were obtained with an Agilent Spectrometer (DD2-700) equipped with a cryoprobe (5 mm H/FCN) at 699.02 MHz 1H frequency (658.40MHz for 19F). All experiments were run at room temperature (25+/-0.1 C). Relaxation delay was set to 35 seconds for quantitative measurement, determined by T1 relaxation time measurements. Each spectrum was collected with 16 scans for adequate signal to noise ratios.

**Photoluminescence peak tracking (PL)**

PL peak tracking was conducted with a Renishaw inVia Raman spectrometer utilizing a 532 nm continuous wave (CW) laser excitation source. The samples were focused on and subsequently continuously illuminated with either 1 or 10 sun intensity laser power for 12 hours or 20 minutes, respectively. A series of 60 and 80 evenly spaced PL acquisitions were made over the 12 hour and 20-minute periods. The acquisitions were analyzed using MATLAB.

**Time-resolved photoluminescence (TRPL)**

A Horiba Fluorolog time-correlated single-photon-counting system with photomultiplier tube detectors was used for time resolved PL measurements. The laser was directed onto the film surface of the perovskite sample. A DeltaDiode laser diode ($\lambda$ = 504 nm) was used for the excitation source with a power density of 2 mW/cm$^2$, corresponding to a low-level injection condition of ~10$^{15}$ cm$^{-3}$. PL decay curves were fitted with bi-exponential function to obtain a fast and slow decay.

**PLQY measurements**

The excitation source was an unfocused beam of a 442 nm continuous-wave diode laser. Photoluminescence was collected using an integrating sphere with a pre-calibrated fiber coupled to a spectrometer (Ocean Optics QE Pro) with an intensity of ~100 mW cm$^{-2}$. PLQY values were calculated by $PLQY = \frac{P_S}{P_{Ex}*A}$ and $A = 1 - \frac{P_L}{P_{Ex}}$. $P_S$ is the integrated photon count of sample emission upon laser excitation; $P_{Ex}$ is the integrated photon count of the excitation laser when the sample is removed from integrating sphere, and $P_L$ is the integrated photon count of excitation laser when sample is mounted in the integrating sphere and hit by the beam.

**Density Functional Theory (DFT) calculations**

All mixed halide perovskites (formula ABX$_3$ in a cubic structure containing various A-cations (A=Cs, FA, MA, EA, DMA, GUA), B-cations (B=Pb, Ba, Sn, Sr), and X-anions (X= I, Br, Cl, $BH_4$, $BF_4$, CN, SCN)) are modeled as a statistical ensemble of independent configurations. The following equation was used to evaluate the Helmholtz free energy:

$$\Delta F(x,T)=\Delta U(x,T)-T\Delta S(x,T) \quad (1)$$

Where $\Delta U(x,T)$ is the internal energy of the alloy, and $\Delta S(x,T)$ is the configurational entropy. To obtain these terms, the total energy $E_i$ of each independent configuration $i$ is calculated by DFT. The total number of configurations for these systems are $2^6 = 64$ (6 halide atoms in the supercell) for $APb(I_{1-x}Br_x)_3$ (A=Cs, FA, MA, EA, DMA, GUA), $CsB(I_{1-x}Br_x)_3$ (B=Pb, Ba, Sn, Sr), and $FAPb(I_{1-x}X_x)_3$ (X= Cl, $BH_4$, $BF_4$, CN, SCN), and $2^{12} = 4096$ (12 halide atoms in the cell) for $MAPb(I_{1-x}Br_x)_3$ (in tetragonal structures) and $FA_{0.75}Cs_{0.25}Pb(I_{1-x}Br_x)_3$, and $2^{11} = 2048$ (11 halide atoms and 1 $BF_4$ "atom" in the supercell) for $FA_{0.75}Cs_{0.25}Pb(I_{0.92-x}Br_x(BF_4)_{0.08})_3$. Then, the total independent configurations are calculated by using the SOD code [40]. By considering the symmetry, there are 21 independent configurations for $APb(I_{1-x}Br_x)_3$ (A=Cs, FA, MA, EA, DMA, GUA), $CsB(I_{1-x}Br_x)_3$ (B=Pb, Ba, Sn, Sr), and $FAPb(I_{1-x}X_x)_3$ (X= Cl, BH4, BF4, CN, SCN), 265 independent configurations for $MAPb(I_{1-x}Br_x)_3$ (in tetragonal structures), 270 independent configurations for $FA_{0.75}Cs_{0.25}Pb(I_{1-x}Br_x)_3$, and 2008 independent configurations for $FA_{0.75}Cs_{0.25}Pb(I_{0.92-x}Br_x(BF_4)_{0.08})_3$. Then the GQCA model is applied by using the code found in reference 21. The additional code used for the light effects and associated phase diagram plotting can be found in the GitHub website. (https://github.com/ericzhut/Model-for-light-induced-phase-segregations)

All DFT calculations were performed using the FHI-aims [41-43] all-electron code. The default numerical settings, referred to as "intermediate" in FHI-aims were used. Local minimum-energy geometries of the Born-Oppenheimer surface were obtained with residual total energy gradients below $1\times10^{-2}$ eV/Å for atomic positions using PBE-GGA functional within the vdW correction following the TS approach (PBE+TS). Different Γ-centered k-space were used for different compounds: $4 \times 6 \times 6$ for $APb(I_{1-x}Br_x)_3$ (A=Cs, FA, MA, EA, DMA, GUA), $4 \times 8 \times 8$ for $CsB(I_{1-x}Br_x)_3$ (B=Pb, Ba, Sn, Sr), $FAPb(I_{1-x}X_x)_3$ (X= Cl, BH4, BF4, CN, SCN), $6 \times 4 \times 4$ for $MAPb(I_{1-x}Br_x)_3$ (in tetragonal structures), and $6 \times 3 \times 3$ for $FA_{0.75}Cs_{0.25}Pb(I_{1-x}Br_x)_3$ and $FA_{0.75}Cs_{0.25}Pb(I_{0.92-x}Br_x(BF_4)_{0.08})_3$. The PBE-GGA functional within the spin-coupling effects is used to obtain the band structures and associated band gaps.

**Supplementary Text 1: Interpretation of the general parameter η in the proposed model**

First, we consider an illuminated area of perovskite film (A), the volume of the perovskite affected is A· $l$ where $l$ can be an arbitrary length but in our case is the film thickness.

To produce an expression for photon energy density in the steady state, we consider the dissipation of this energy within the perovskite through ion migration. The time constant of this energy transfer can be evaluated according to the ion diffusion decay time, $\tau_{ion}$, which can be evaluated through halide transport in our region of interest by the following equation:

$$l = \sqrt{D_{ion}\tau_{ion}}$$

Where the $D_{ion}$ is the diffusion coefficient of the ions (e.g., halides in the mixed halide perovskite). Then, the steady state number of photon energy units, N[photon], in the time $\tau_{ion}$ is evaluated by the following equation:

$$N[photon] = \Phi_{photon} \cdot A \cdot (1 - e^{-\alpha \cdot l}) \cdot \tau_{ion} = \Phi_{photon} \cdot A \cdot (1 - e^{-\alpha \cdot l}) \cdot \frac{l^2}{D_{ion}}$$

Where $\Phi_{photon}$ is the photon flux number according to different light intensity, α is the absorption coefficient, and the penetration length is estimated as the film thickness $l$.

On the other hand, the total number of perovskite units, N[PV] can be evaluated through the following equation:

$$N[PV] = \frac{A \cdot l}{\Omega_{cell}}$$

Where $\Omega_{cell}$ is perovskite unit cell volume.

Then the general parameter describing photon energy density, η in the proposed theory, can be represented through the following equation:

$$\eta = \frac{N[Photon]}{N[PV]} = \frac{\Omega_{cell} \cdot \Phi_{photon} \cdot (1 - e^{-\alpha \cdot l}) \cdot \tau_{ion}}{l}$$

$$= \frac{\Omega_{cell} \cdot \Phi_{photon} \cdot (1 - e^{-\alpha \cdot l}) \cdot l}{D_{ion}}$$

For different mixed halide perovskites, the band gap will be different. This will further result in different $D_{ion}$, $\Omega_{cell}$, $\alpha$, $l$, and $\Phi_{photon}$ for 1-sun intensities. In this paper, all the parameters used to relate $\eta$ to light intensities are summarized in Table S7. Following the above equation, for the MAPbI$_3$ system, we calculate the experimentally used 1 sun laser intensity to η= 1/(6.42±1.07) based on the bandgap of 40% Br, ~ 1.75 eV, with a $D_{ion}$, ~ 6 ± 1 × 10-9 cm$^2$s$^{-1}$ ; for (FA$_{0.75}$Cs$_{0.25}$)Pb(I$_{0.6}$Br$_{0.4}$)$_3$ and the BF$_4^-$ doped composition, we calculate the experimentally used 1 sun laser intensity to $\eta$ = 1/(7.0±1.2) based on the bandgap of 40%Br, ~1.78 eV, with the same $D_{ion}$, ~ 6 ± 1 × 10$^{-9}$ cm$^2$s$^{-1}$. We note that this interpretation is rough because the diffusion time $\tau_{ion}$ is difficult to determine experimentally, and discrepancy in ion diffusion coefficient will contribute to the uncertainty in interpretating $\eta$ values. Furthermore, the photon flux (among other parameters) of 1 sun laser intensity also changes across perovskite alloy compositions.

However, the proportional relationship between $\eta$ and $\tau_{ion}$ has an experimental basis: when one uses a pulsed laser with a low repetition rate, halide segregation does not occur, accounted for by the fact that there is enough time for the energy of each pulse to dissipate effectively. [44,45] These studies suggest that segregation begins for repetition rates above 1-5kHz, or time between pulses faster than ~200μs. Using our calculated decay constant, $\tau_{ion}$~15 μs, and assuming an exponential decay, the lattice would retain only 0.0002% of the original pulse energy after 200 μs: thus, the energy is effectively dissipated. Though indirect evidence, this demonstrates that the range of our equivalence between $\eta$ and Φ$_{photon}$ is reasonable.

**Supplementary Text 2: Experimental evidence of $BF_4^-$ incorporation within the perovskite lattice**

Although evidence of $BF_4^-$ incorporation has been observed in certain perovskite compositions [46,47], it remains debated in others [48]. We thus set out to conduct our own experiments to confidently infer $BF_4^-$ incorporation. We prepared thin film ($FA_{0.83}Cs_{0.17}Pb(I_{0.6}Br_{0.4})_3$) and single crystal ($FAPbBr_3$) perovskites with $FABF_4$ in the precursor solution (details in above methods). As seen in prior reports [46] the lattice expands (XRD of both thin-film and single crystal samples) with increasing $BF_4^-$ content (Fig. S9, S16), and a small shift in optical bandgap (Fig. S10) and PL emission wavelength occurs (Fig. S14), both of which suggest alloying, but could (absent other evidence) be explained by passivation or redistribution of perovskite ions. To further understand alloying potential more deeply, we grew single crystal samples ranging from 1% mol to 10% mol. excess $FABF_4$ in the precursor solutions and washed the samples thoroughly using acetonitrile (of which $BF_4^-$ is soluble in) to assure that subsequent measurements (NMR, XPS, UV-VIS, and PL) are not simply detecting residual $BF_4^-$ on the surface of the crystal. We dissolved single crystals and used NMR to verify the presence of $BF_4^-$ (Fig. S17), and a combination of XPS and sputtering XPS to confirm that $BF_4^-$ was in the interior of single crystals, not only the surface (Fig. S12c), and quantified the ratio of Br to $BF_4^-$ in crystals grown with different $BF_4^-$ percentages to be a roughly linear function between 0.1% - 0.5% (Fig S18 and Table S2).

**Supplementary Text 3: Device discussion**

We fabricated perovskite solar cells using an (ITO/NiOx/2PACz/perovskite/C$_{60}$/BCP/Au) architecture, the details of which are in the methods (solar cell fabrication). BF$_4^-$ inclusion improved power conversion efficiency (PCE), efficiency in the main text, via increases to fill factor and open-circuit voltage (V$_{OC}$) (Fig. 4a) from current-voltage (J-V) scans of a sample of 15 devices (Fig. S19). This improvement is consistent with increased PL lifetimes and PL intensity (Fig. S13-S14 and Table S3) [46,49]. We tracked unencapsulated solar cell efficiency at the maximum power point under 1 sun illumination to best match illumination conditions under which the PL tracking was performed.

**Supplementary Figures:**

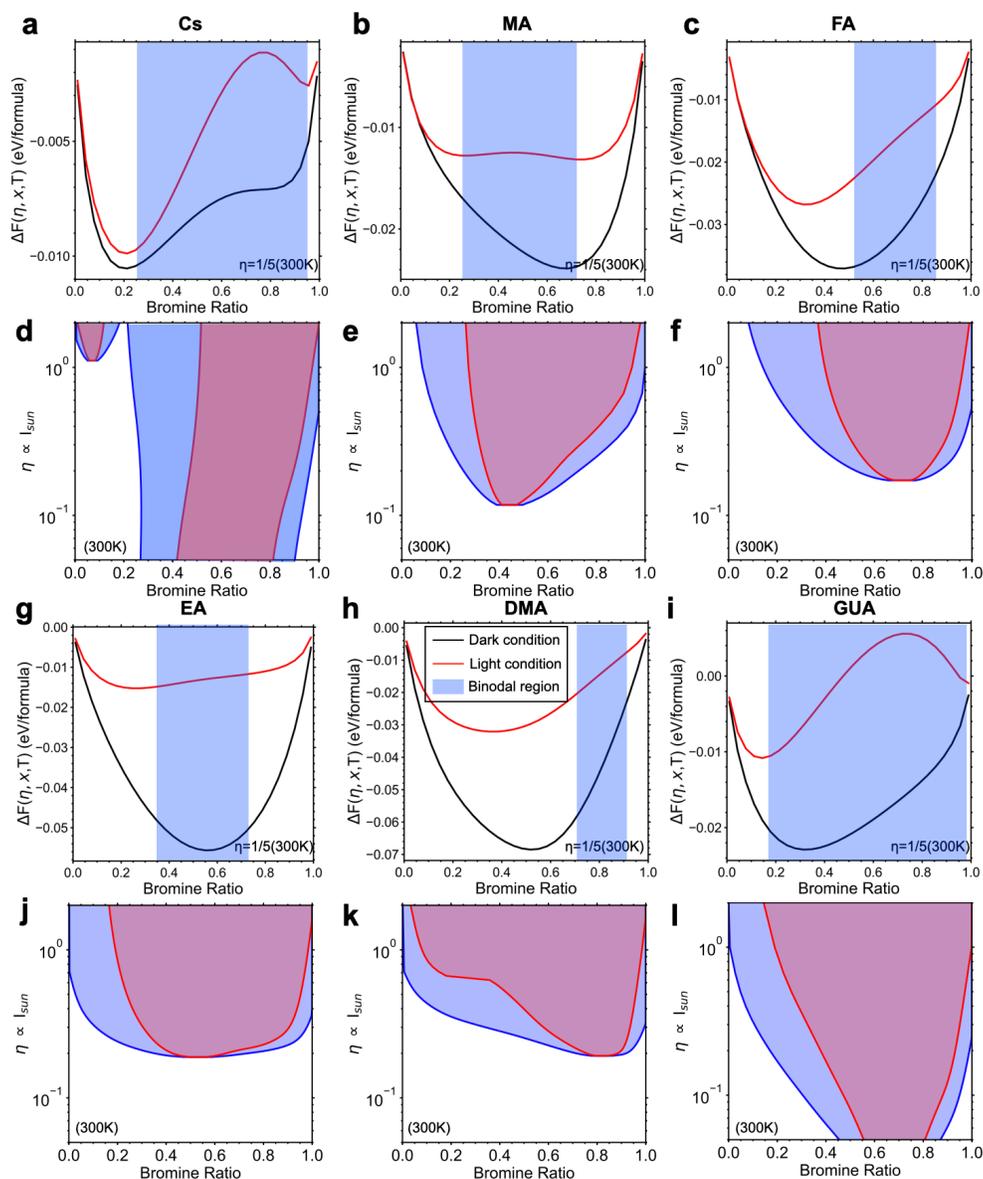

**Figure S1.** The ΔF(η, *x*, T) calculated for different mixed halide perovskites (APb(I$_{1-x}$Br$_x$)$_3$) in cubic structures at T=300K, and η=1/5, and the predicted phase diagrams along with different light intensities at 300K, in which A = a,d) Cs; b,e) MA; c,f) FA; g,j) EA; h,k) DMA; i,l) GUA.

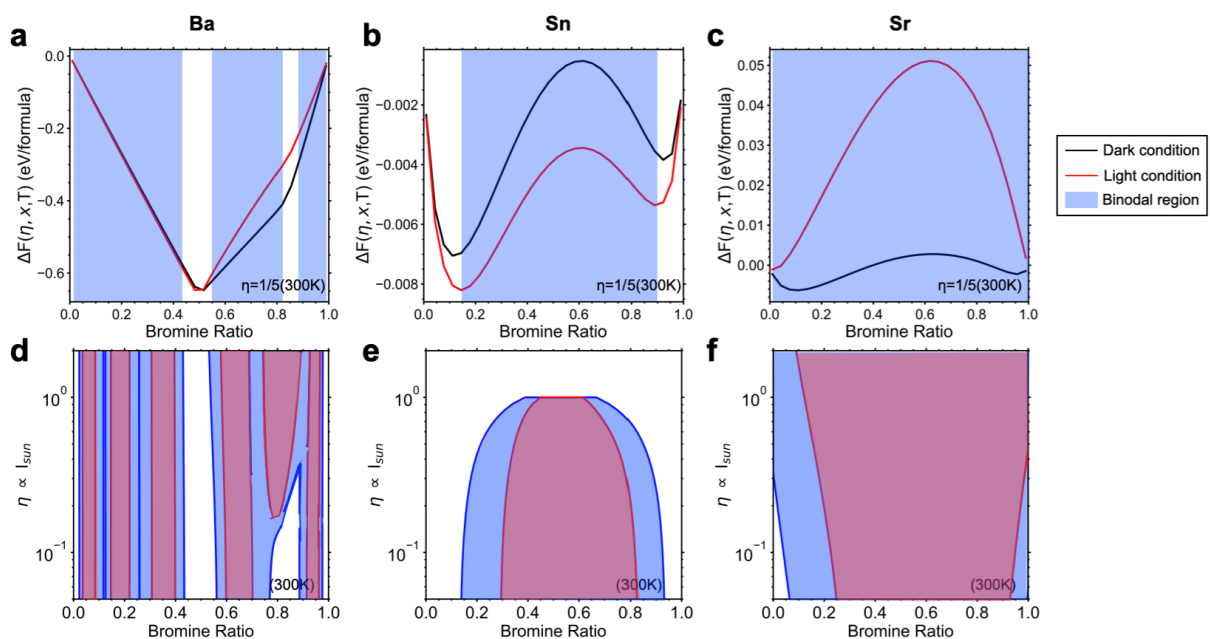

**Figure S2**. a-c) The $\Delta F(\eta, x, T)$ calculated for different mixed halide perovskites (CsB(I$_{1-x}$Br$_x$)$_3$) in cubic structures at 300K, $\eta=1/5$, and the predicted phase diagrams along with different light intensities at 300K, in which B = a,d) Ba; b,e) Sn; and c,f) Sr.

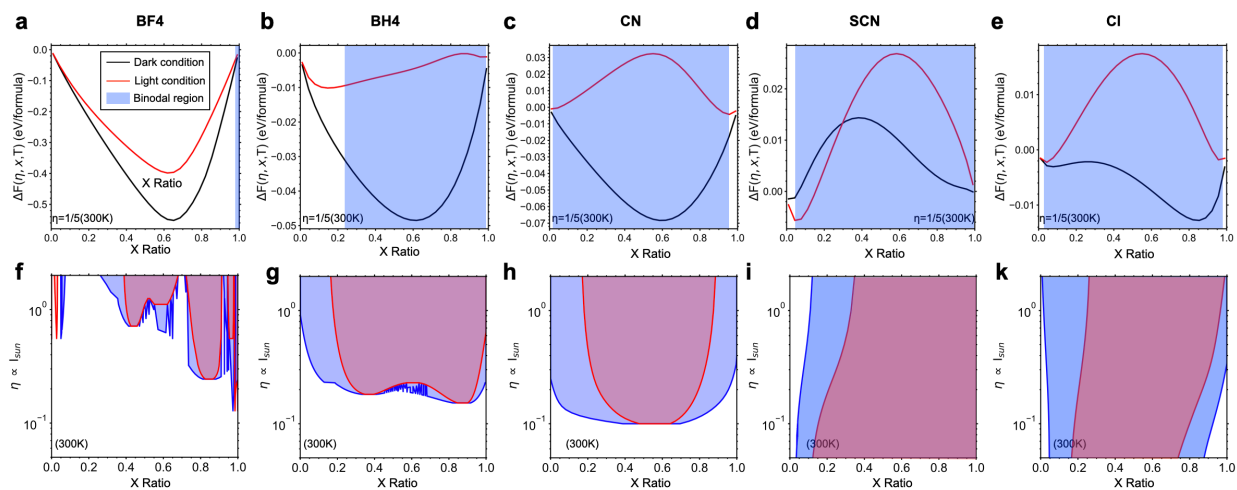

**Figure S3**. a-e) The $\Delta F(\eta, x, T)$ calculated for different mixed halide perovskites (FAPb(I$_{1-x}$(X)$_x$)$_3$) in cubic structures at 300K, and $\eta=1/5$, in which X = BF$_4$, BH$_4$, CN, SCN, Cl. f-k) The predicted phase diagrams along different light intensities at 300K. The binodal and spinodal lines and regions are shown in blue and red, respectively.

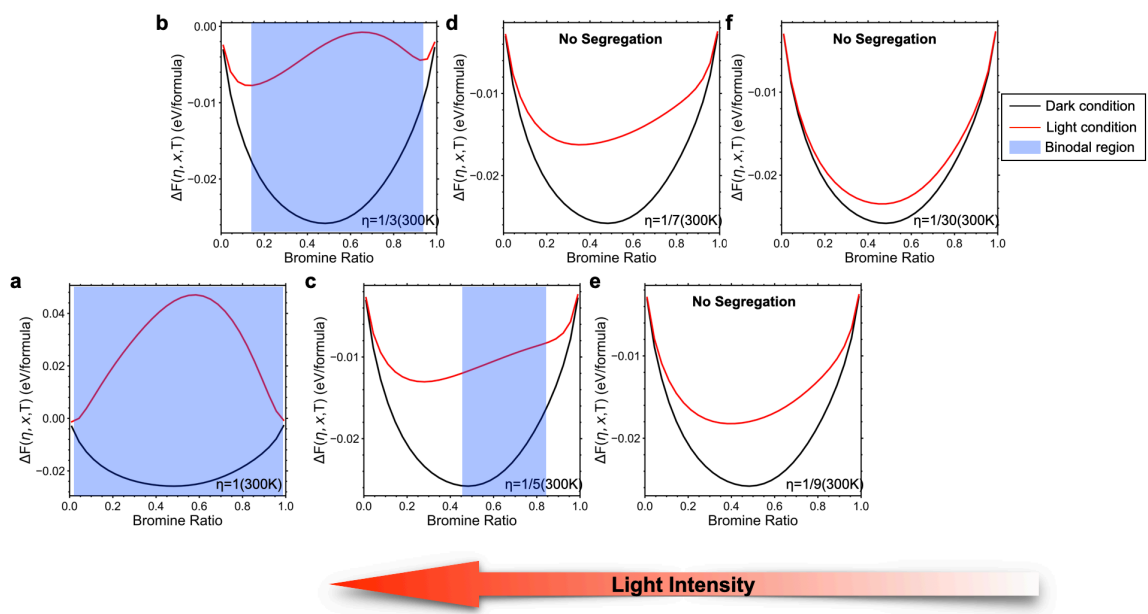

**Figure S4**. The predicted Helmholtz free energy variation plots with different light intensities at 300 K for MAPb($I_{1-x}Br_x$)$_3$. The binodal region is shown as a blue shaded area.

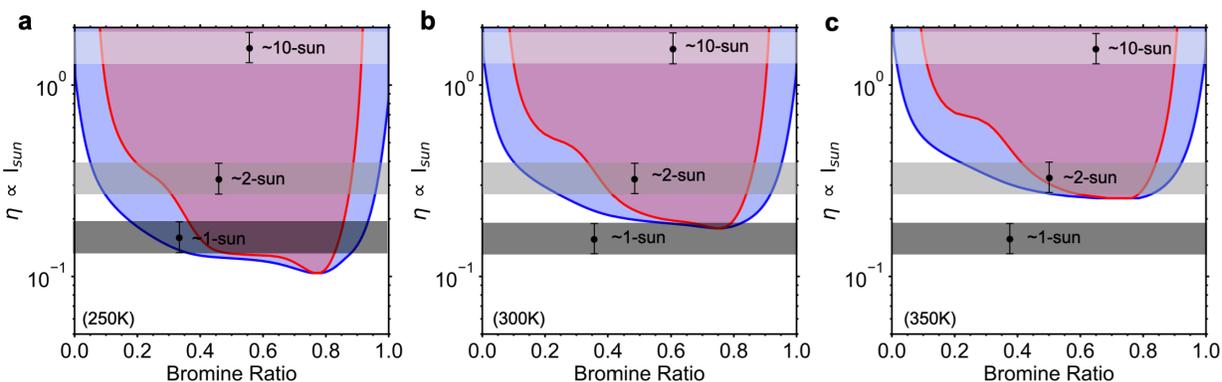

**Figure S5**. The predicted phase diagrams along with different light intensities at different temperatures: a) 250K, b) 300K, and c) 350K for MAPb($I_{1-x}Br_x$)$_3$. The binodal and spinodal lines and regions are shown in blue and red, respectively. The estimated light intensities 1-sun, 2-sun, 10-sun are illustrated by greyscale shaded areas.

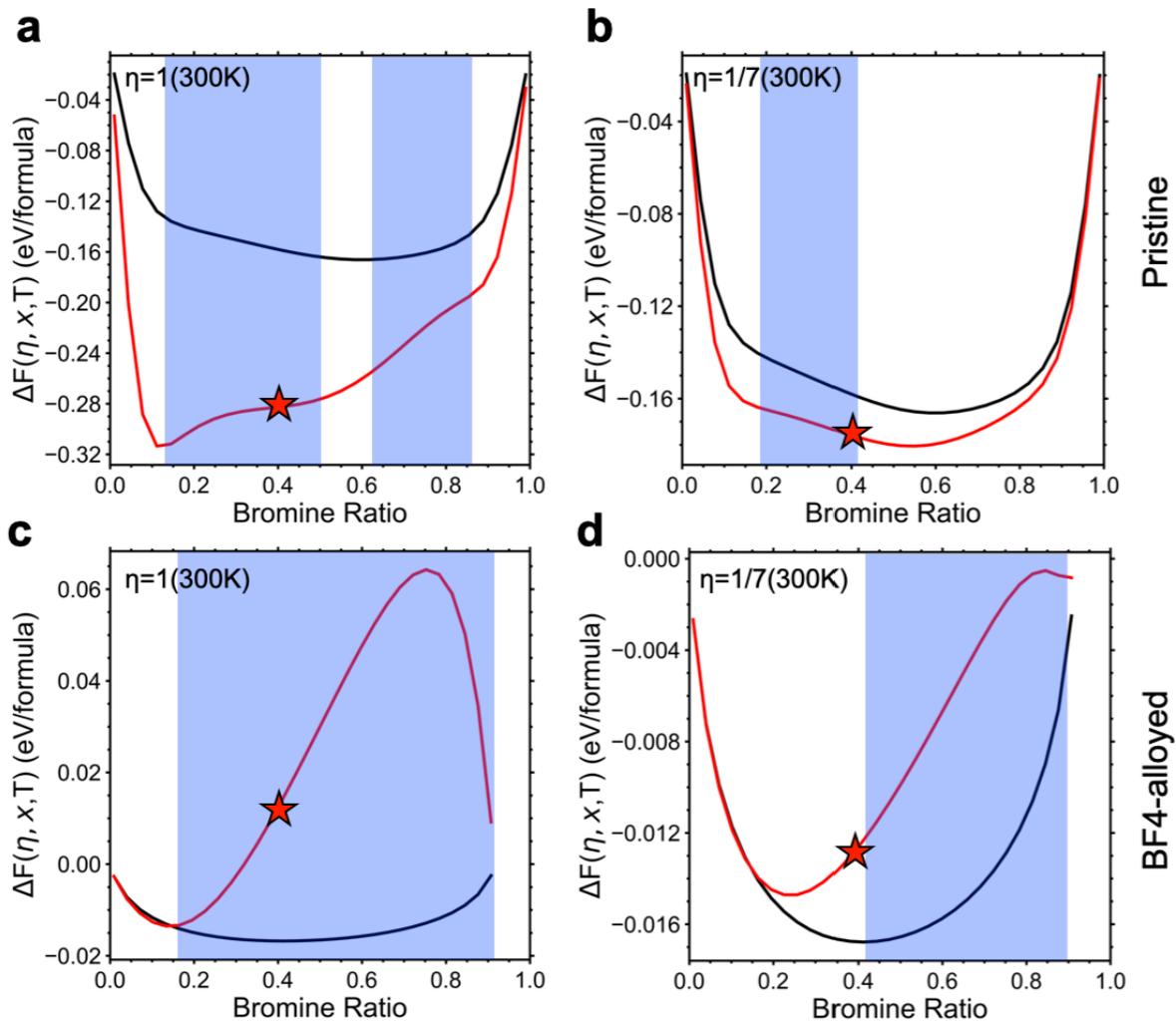

**Figure S6**. The predicted Helmholtz free energy variation plots within different light intensities at 300 K for a,b) $FA_{0.75}Cs_{0.25}Pb(I_{1-x}Br_x)_3$ (Pristine) and c,d) $FA_{0.75}Cs_{0.25}Pb(I_{0.92-x}Br_x(BF_4)_{0.08})_3$ ($BF_4^-$-alloyed). The target concentrations at 40% Br concentration are labeled as the red star. The binodal region is shown as a blue shaded area.

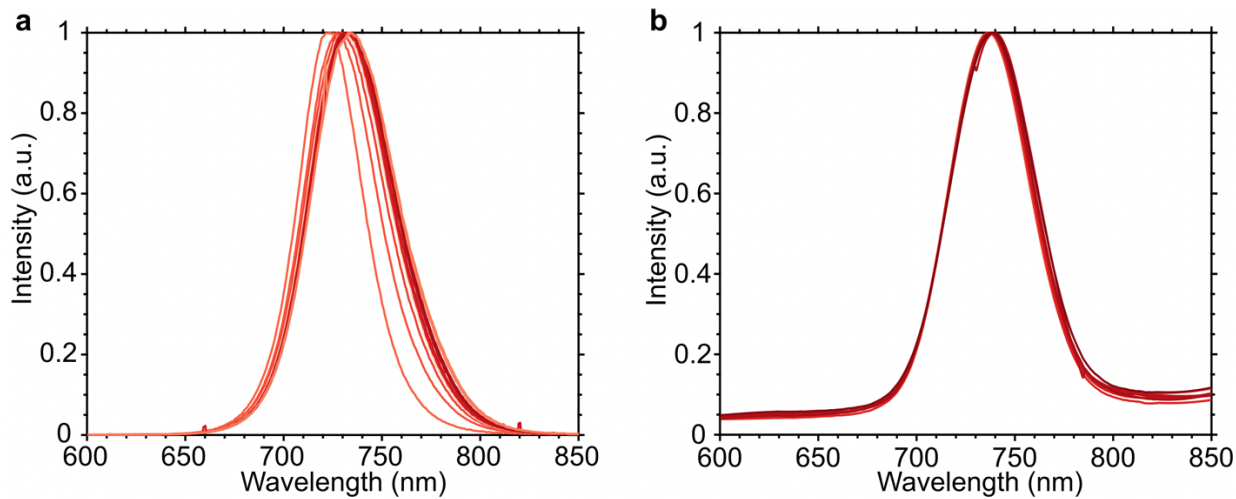

**Figure S7**. PL peak tracking of 30% Br films at 1 sun intensity over a period of 12 hours. a) 30% Br control perovskite film. b) 1% $BF_4^-$ containing perovskite film.

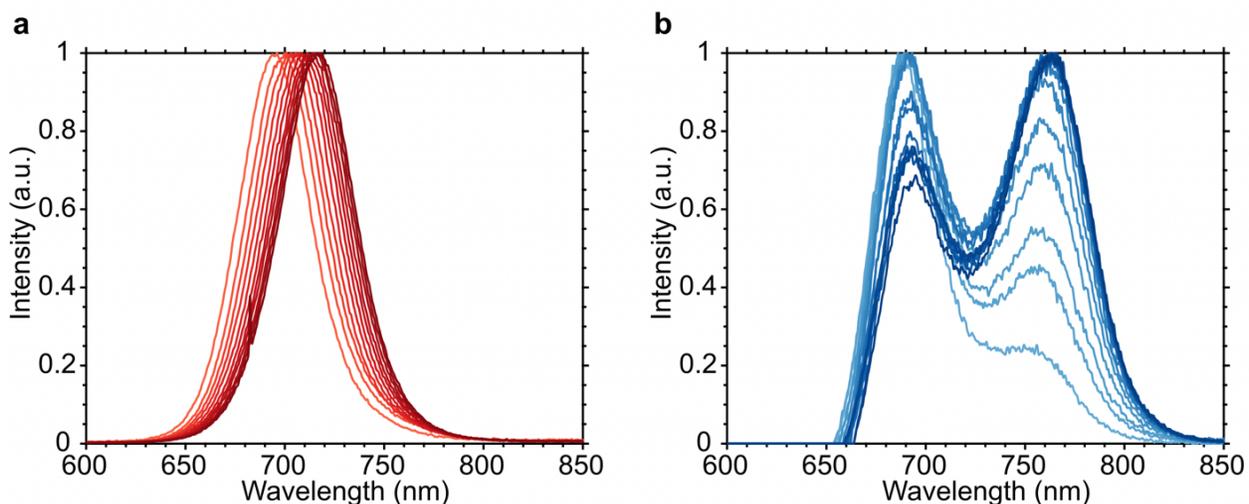

**Figure S8**. PL peak tracking of 40% Br films containing 2 mole percent chlorine. a) Tracking at 1 sun intensity for a period of 12 hours. b) Tracking at 10 sun intensity over a period of 20 minutes.

|  | Control Peak Shift (nm) | With $BF_4^-$, Peak Shift (nm) | With Cl, Peak Shift (nm) |
|---|---|---|---|
| 30% Bromine | 10 | 3 | - |
| 40% Bromine, 1 sun | 85 | 18 | 35 |
| 40% Bromine, 10 suns | 73 | 68 | 76 |

**Table S1**. Tabulated peak shifts for PL peak tracking of perovskite thin films with and without additives at various intensities.

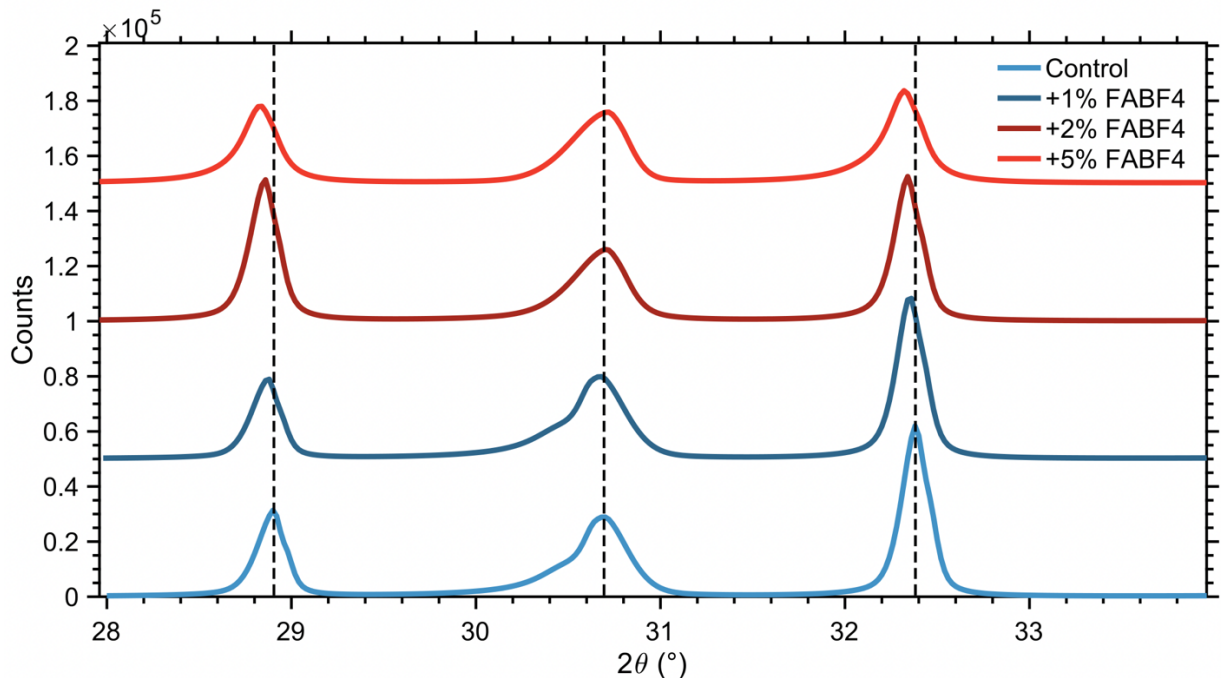

**Figure S9**. Thin-Film XRD of 40% Br perovskite thin films with various $BF_4^-$ loading percentages.

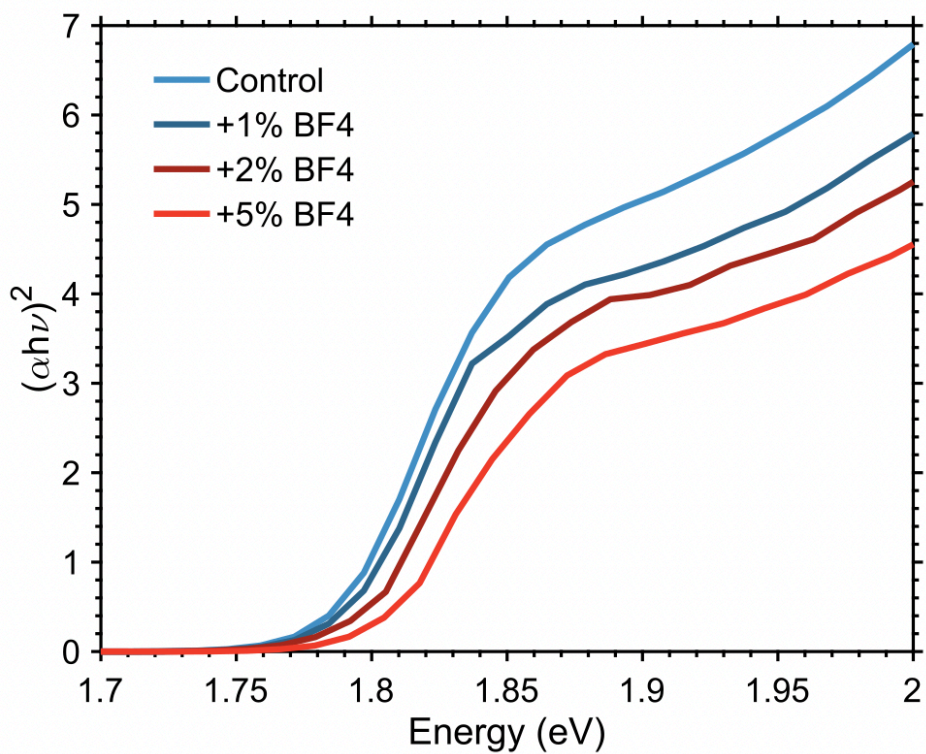

**Figure S10**. Ultraviolet-visible spectroscopy of perovskite thin films with various $BF_4^-$ loading percentages.

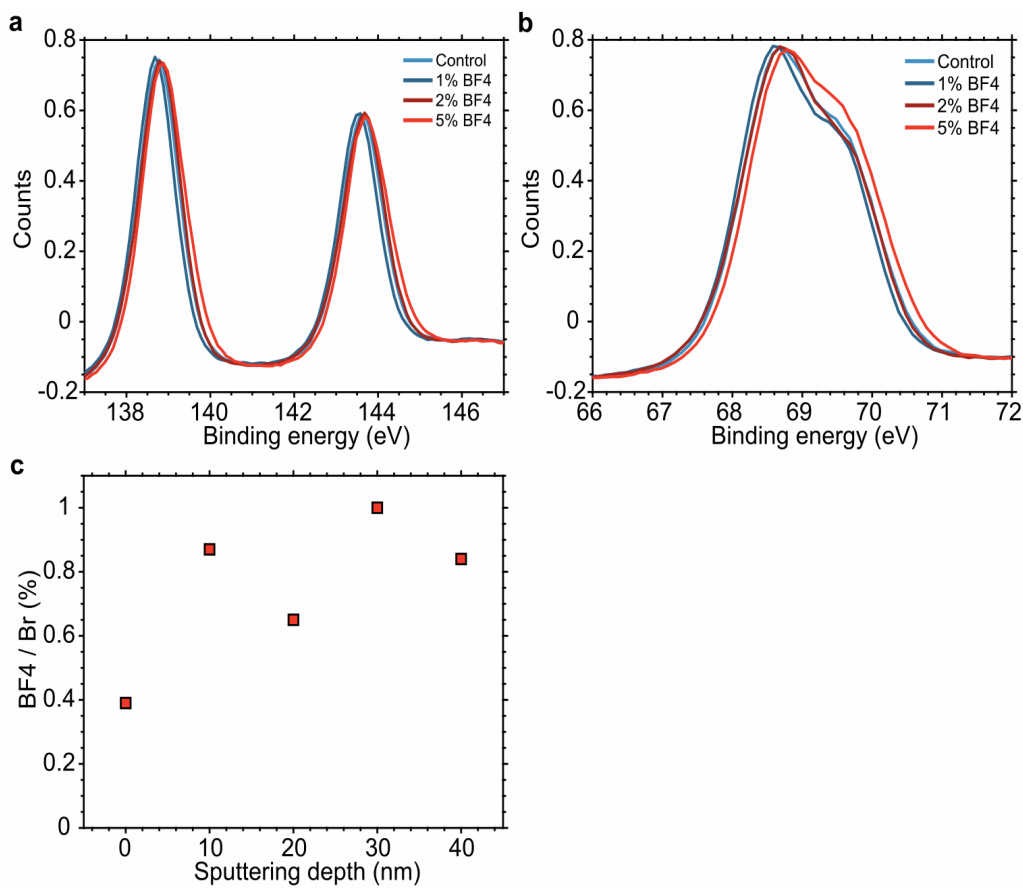

**Figure S11**. XPS measurements of FAPbBr$_3$ perovskite single crystals with various BF$_4^-$ loading percentages. a) Pb 4f 5/2 and 7/2 shift to a lower binding energy with increasing BF$_4^-$ percentage. b) Br 3d c) Sputtering XPS of BF$_4^-$ containing single crystal emphasizing the presence of BF$_4$ at a depth into the single crystal.

| Crystal | BF$_4^-$ / Br (%) |
| --- | --- |
| Control | 0 |
| 1% BF$_4^-$ | 0.17 |
| 2% BF$_4^-$ | 0.30 |
| 5% BF$_4^-$ | 0.57 |

**Table S2**. Quantification of BF$_4^-$ present in the FAPbBr$_3$ single crystals with respect to amount of BF$_4^-$ added in the crystal growth solution. BF$_4^-$ / Br (%) was acquired by dividing the peak area of the fluorine 1s signal by 4 (4 fluorine atoms per tetrafluoroborate anion), divided by the peak area of the bromine 3d signal.

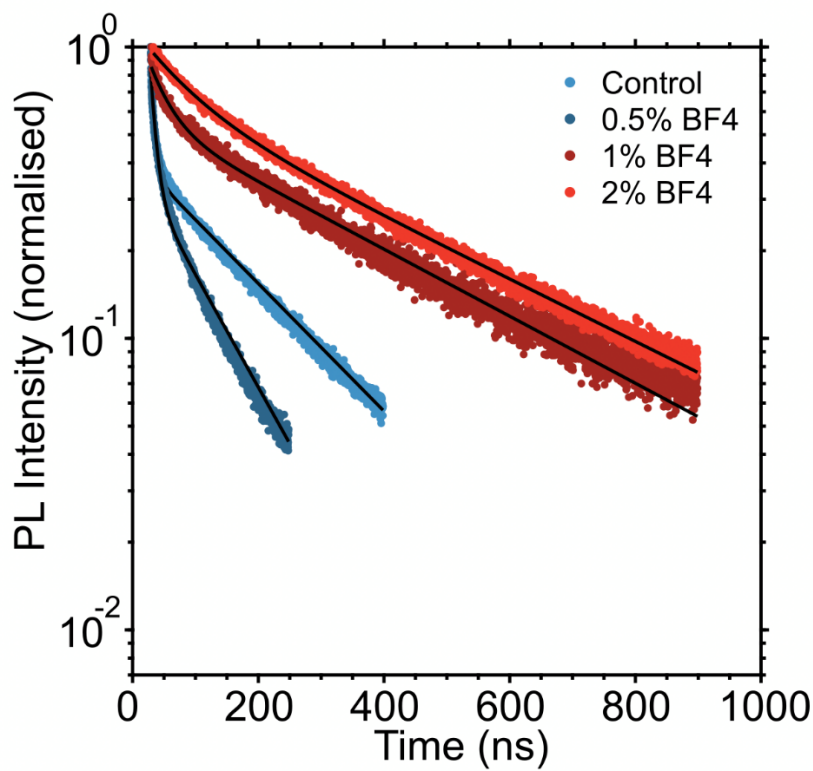

**Figure S12.** TRPL measurements of 40% Br perovskite thin films containing various $BF_4^-$ loading percentages.

|  | $t_1$ (ns) | $t_2$ (ns) |
|---|---|---|
| Control | 6 | 198 |
| +0.5% $BF_4^-$ | 9 | 112 |
| +1% $BF_4^-$ | 33 | 378 |
| +2% $BF_4^-$ | 78 | 406 |

**Table S3.** Tabulation of fast ($t_1$) and slow ($t_2$) lifetimes for 40% Br perovskite thin films at various $BF_4^-$ loading percentages.

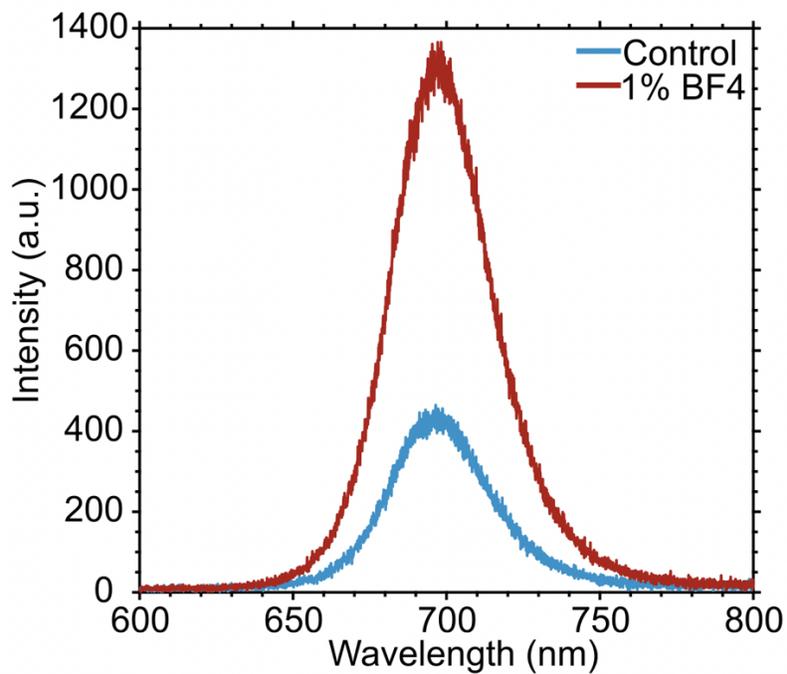

**Figure S13**. First scan PL intensities for 40% Br containing perovskite with and without $BF_4^-$.

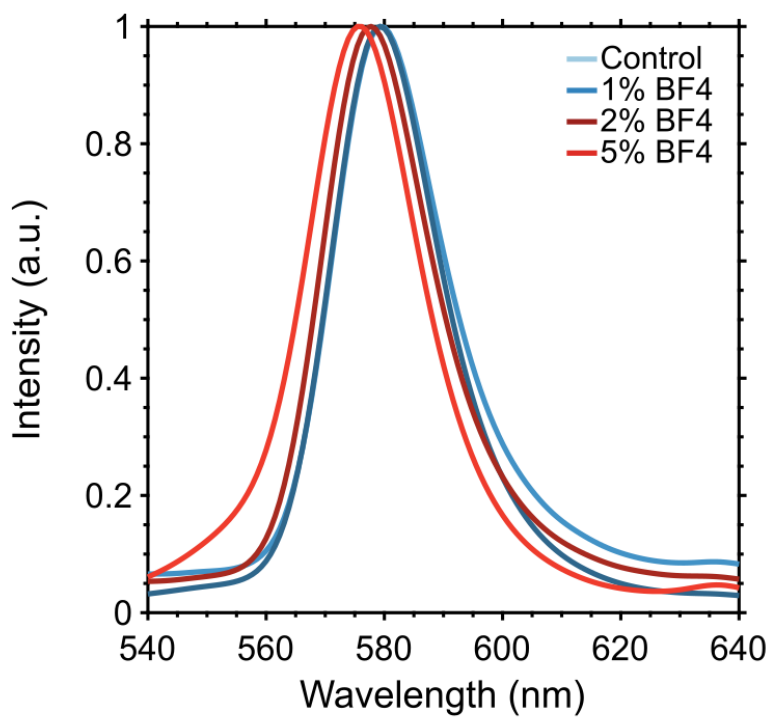

**Figure S14**. PL measurements of crushed $FAPbBr_3$ single crystals with various $BF_4^-$ percentages.

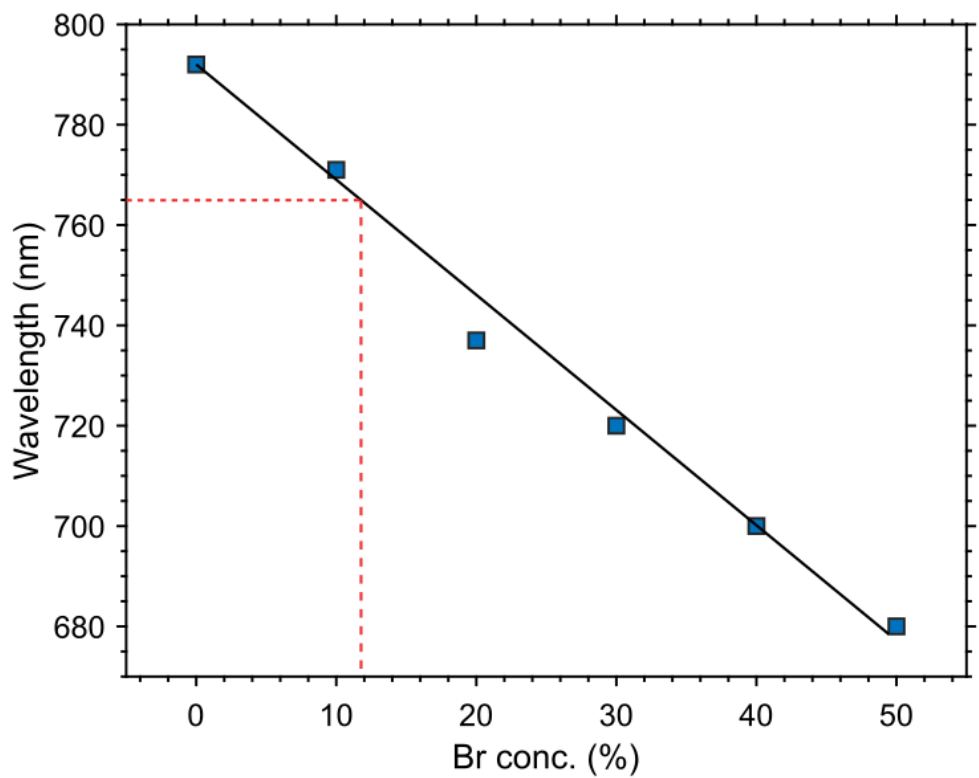

**Figure S15**. PL emission as a function of Br% (x) in $FA_{0.83}Cs_{0.17}Pb(I_{1-x}Br_x)_3$. Red lines indicate the 765 nm PL emission which corresponds to ~12% Br, the iodine rich segregated phase.

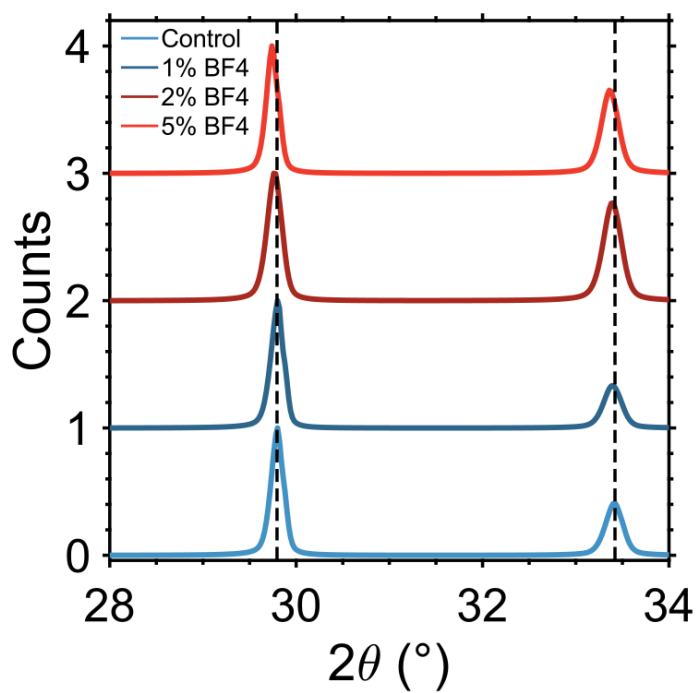

**Figure S16.** Crushed single crystal (powder) XRD spectra of $FAPbBr_3$ single crystals grown with different $BF_4^-$ percentages.

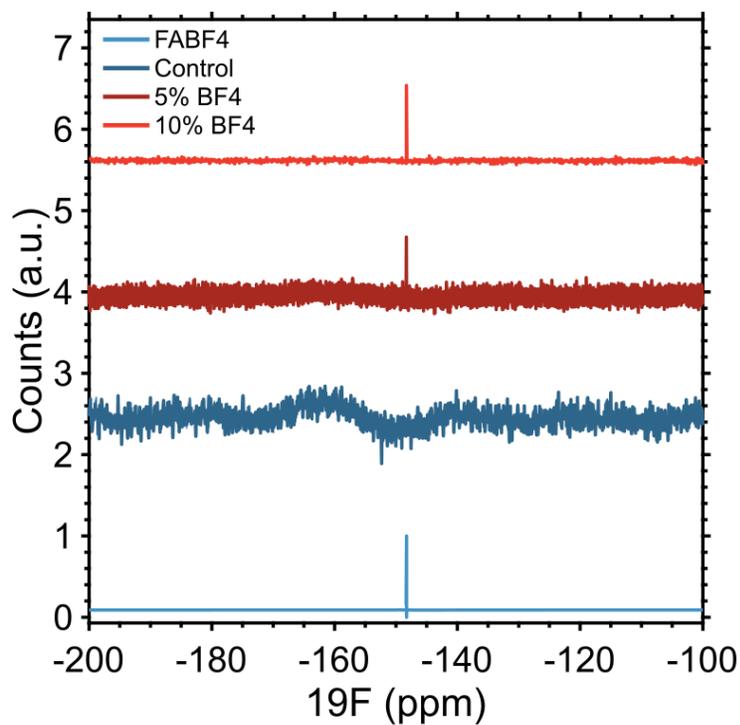

**Figure S17.** Fluorine solution NMR from washed FAPbBr$_3$ single crystals grown with different BF$_4^-$ percentages.

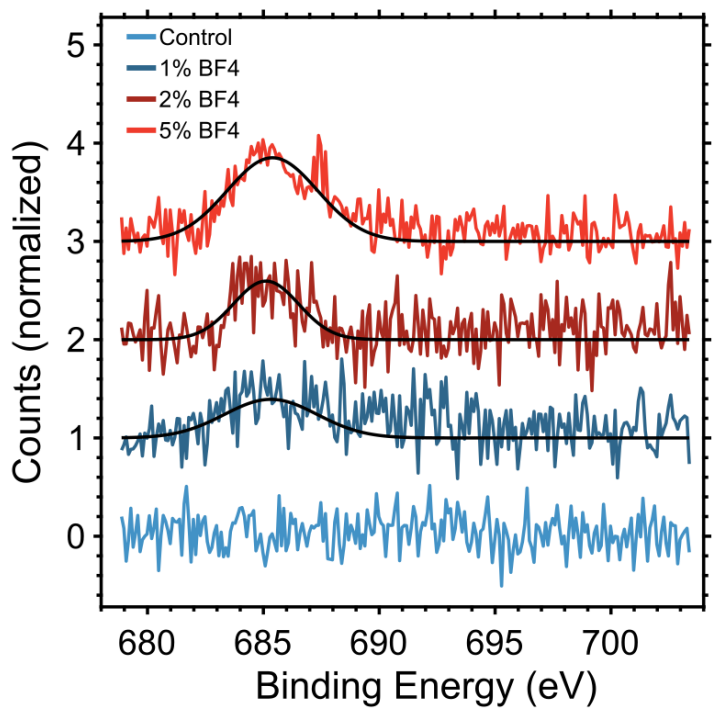

**Figure S18.** XPS data from washed FAPbBr$_3$ single crystals grown with different BF$_4^-$ percentages.

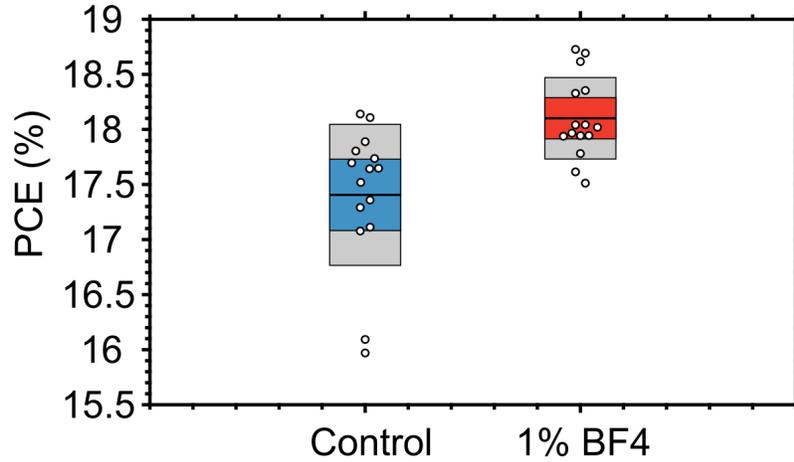

**Figure S19.** Box plots demonstrating the spread of PCE values for control and $BF_4^-$ perovskite solar cells.

| $MAPb(I,Br)_3$ | | $FA_{0.75}Cs_{0.25})Pb(I_{0.6}Br_{0.4})_3$ and $BF_4^-$ doped one | |
|---|---|---|---|
| flux$_{photon}$ (at 1-sun)[52] | 1.47e+17 cm$^{-2}$s$^{-1}$ | flux$_{photon}$ (at 1-sun) | 1.37e+17 cm$^{-2}$s$^{-1}$ |
| $D_{ion}$[50,51] | $6 \pm 1 \times 10^{-9}$ cm$^2$ s$^{-1}$ | $D_{ion}$ | $6 \pm 1 \times 10^{-9}$ cm$^2$ s$^{-1}$ |
| $l$ | 288 nm | $l$ | 288 nm |
| $\Omega_{PV}$ | 249 Å$^3$ | $\Omega_{PV}$ | 234 Å$^3$ |
| $\alpha$ | 1.5e+05 cm$^{-1}$ | $\alpha$ | 7.03e+04 cm$^{-1}$ |

**Table S4**. The parameters used for interpretation of η to real light intensities.

| Ref [4] | | | |
|---|---|---|---|
| Compounds | 60% Br , 1.85 eV | Laser | 457 nm, 50 mW cm$^{-2}$ |
| Experimental observed Iodine-rich phase | 20% Br | Theory predicted Iodine-rich phase | Close to stable |
| | | | |
| Ref[30] | | | |
| Compounds | 26% Br, 1.65 eV | Laser | 532nm, 3-suns |
| Experimental observed Iodine-rich phase | ~13% Br | Theory predicted Iodine-rich phase | ~5% ~ 9% |
| | | | |
| Ref[31] | | | |
| Compounds | 48% Br, 1.80 eV | Laser | 1-suns |
| Experimental observed Iodine-rich phase | ~45% Br | Theory predicted Iodine-rich phase | ~53% |
| | | | |
| Ref[29] | | | |
| Compounds | 30% Br, | Laser | 1.7-suns |
| Experimental observed Iodine-rich phase | ~20% Br | Theory predicted Iodine-rich phase | ~15-34% |

**Table S5**. The comparisons between theory predicted and experimentally observed iodine rich phase formation after illumination in MAPb( I$_{1-x}$Br$_x$)$_3$.